# Ethaline deep eutectic solvent under nanoconfinement: Unveiling structural and dynamical changes

Mohammad Nadim Kamar[a], Armin Mozhdehei[a], Ronan Lefort[a], Alain Moréac[a], Jacques Ollivier[b], Markus Appel[b], Viviana Cristiglio[b] and Denis Morineau[a*]

[a]Institute of Physics of Rennes, CNRS-University of Rennes, UMR 6251, F-35042 Rennes, France

[b]Institut Laue-Langevin, 71 avenue des Martyrs, F-38000 Grenoble, France

* denis.morineau@univ-rennes.fr

# ABSTRACT

Hybrid nanomaterials incorporating deep eutectic solvents (DES) in porous hosts or at solid interfaces are gaining increasing attention for their potential interest across a wide range of applications. Under these conditions, the performances of DESs may be influenced by interfacial effects and spatial restrictions. In this study, we examined the effects of nanoconfinement on both the structure and molecular dynamics of the prototypical DES ethaline (a mixture of choline chloride and ethylene glycol) when confined within the cylindrical mesopores of SBA-15 ($D_p \approx$ 8.1 nm) and MCM-41 ($D_p \approx$ 3.5 nm) silicas, using neutron diffraction and quasielastic neutron scattering. It demonstrates that ethaline remains structurally homogeneous under confinement, showing no evidence of core–shell segregation within the pore cross-section. The molecular dynamics of the confined ethaline preserve the key characteristics observed in its bulk state. Translational diffusion follows a jump-diffusion mechanism, with diffusion coefficients that remain remarkably close to the bulk values, showing only a modest reduction in MCM-41. A more pronounced increase in the residence time $\tau_0$ between translational molecular jumps is observed. It corresponds to roughly a factor of 3-8 under confinement in SBA-15, and reaches up to a tenfold enhancement in MCM-41 relative to bulk. Similarly, the characteristic relaxation time, $\tau_L$, associated with the localized in-cage motion of ethaline, increases by approximately 20% in SBA-15 and up to 50% in MCM-41. However, the molecular trajectories, modeled from the elastic incoherent structure factor, remain largely preserved under confinement, showing only a marginal reduction in intra-basin motional amplitudes.

## INTRODUCTION

Deep eutectic solvents (DES) have been introduced two decades ago as a new class of solvents.[1,2] They are mixtures, usually formed by combining a hydrogen-bond donor (HBD) with a hydrogen-bond acceptor (HBA),[1–5] They exhibit significant melting-point depressions, which advantageously extend their usable temperature range. These non-ideal mixing effects arise from strong specific interactions, such as hydrogen bonds, electrostatics, and supramolecular correlations. DESs are now increasingly recognized as designer solvents offering a promising and sustainable alternative to conventional organic solvents and ionic liquids (ILs) across a wide range of applications.[2,3,6] Many of the potential applications of DESs involve interfacial or nanoconfined systems. These include diverse technological processes, such as $CO_2$ absorption,[7] gas separations,[8] dye-sensitized solar cells,[9,10] electrochemistry,[11,12] and membrane-based processes for liquid separation and for water reclamation via forward osmosis.[13]

The potential for innovation offered by the use of DES in nanoconfined devices underscores the need for fundamental studies to elucidate how confinement affects their intrinsic properties.[14] Indeed, for molecular liquids, many studies have shown that confinement can change important properties of liquids, such as melting and freezing,[15,16] phase transitions,[17,18] and the dynamical slowdown when approaching the glassy state.[19,20] These effects are highly sensitive to the pore size and the nature of the interaction between the liquid and the pore walls. More generally, confinement can modify phase equilibria inside porous media and can even stabilize states or transitions that do not appear in the bulk state. In addition to these global thermodynamic changes, confinement can also create new local structures. For instance, molecules may form structured layers near the pore surface, the hydrogen-bond network can be distorted, and, in the case of mixtures, the local concentration may vary from the pore wall to the center.[21–23] Over the past decades, the crystalline frameworks of zeolites have made them ideal model systems for studying the confinement effects of hydrocarbons in microporous materials.[24,25] In particular, quasielastic neutron scattering experiments, combined with molecular dynamics simulations, have highlighted how guest-host interactions, the geometry of the porous network, and the chemical properties of the guest molecules influence their diffusion behavior.

As with molecular liquids, the impact of confinement on the properties of ILs has been the subject of extensive research over the past few decades.[26] Distinct interfacial structures have been observed at the pore surface, with varying patterns influenced by the nature of the ions, surface chemistry, pore loading, and interactions between the pore surface and the ionic liquid.[26] These nanostructures impact the dynamics of the IL. For instance, a quasielastic neutron scattering (QENS) study reported an unexpected acceleration in dynamics relative to the bulk liquid of $[bmim^+][Tf_2N^-]$ confined in the 8.5 nm diameter pores of a mesoporous carbon matrix.[27,28] This phenomenon was explained by the preferential adsorption of ions onto the pore wall, creating a denser interfacial layer and consequently reducing the density, and increasing the mobility, of ions in the pore's central region. However, this trend is not universal: both enhanced and reduced dynamic properties of confined ILs have been observed, suggesting a strong dependence on the specific molecular characteristics of both the confining matrix and the ionic liquid itself.[26]

Transitioning from pure conventional solvents to binary mixtures, confinement can give rise to remarkable phenomena, including microphase separation in systems that are normally fully miscible in the bulk. This was shown for the tert-butanol/toluene mixture by neutron scattering experiments.[22] When confined within cylindrical silica nanopores, the mixture was shown to adopt a core–shell structure, where one component preferentially accumulates near the pore wall and the other concentrates in the pore center. Mhanna et al.[21] showed that this phenomenon is not limited to the small pores of MCM-41 (~3.6 nm), where it was initially observed, but also extends to the larger pores of SBA-15 (~8 nm). From the dynamical point of view, QENS and calorimetric measurements also revealed strong spatial and dynamical heterogeneities in these confined mixtures, linking the structural segregation to different relaxation processes in the interfacial region and in the pore core.[29,30]

Given these findings, a compelling question arises: what occurs when the confined liquid is not a conventional binary mixture, but rather a deep eutectic solvent (DES), whose structure is governed by strong ionic and hydrogen-bonding interactions?

Our previous bulk study demonstrated that ethaline shows nanometer scale dynamics that differ from those at the macroscopic scale.[31] Different behaviors were observed for similar DESs that belong to the family of choline chloride based systems.[32,33] This length-scale-dependent dynamics observed in DESs are likely related to the specific mesoscopic order they exhibit, particularly in terms of supramolecular association. This assessment is also in line with observations made for a different DES incorporating acetamide and lithium perchlorate. For these systems, a connection was established between the nanoscopic diffusion mechanism, hydrogen bonding, and the formation of molecular complexes.[34–37]

Under confinement, DESs present an even more complex scenario, as multiple competing effects come into play simultaneously: surface affinity to the pore wall, the intrinsic cohesion of the DES network, and finite-size constraints. However, compared to the large amount of work on confined conventional liquids[15,38] and ILs,[26] DESs have been much less studied under confinement.[39] In a study highly relevant to the present work, Malfait et al.[17] explored the impact of confinement on the phase behavior of ethaline, a prototypical DES composed of choline chloride and ethylene glycol. They showed how hydration and glassy dynamics can control thermal transitions in nanoconfined geometry for this system.

In this context, the present study tackles a fundamental question: how do solid interfaces and nanoconfinement change the supramolecular structure and the molecular dynamics of DES? To address this question, we chose model systems that pair SBA-15 and MCM-41 mesostructured porous silicas - both characterized by parallel, ordered cylindrical pores - as host matrices, alongside ethaline, one of the most extensively studied DES in the bulk state to date. High-resolution neutron scattering, complemented by quasielastic neutron scattering using isotopically labeled samples, offers unparalleled insights into both the nanoscale organization and the dynamics of this DES mixture. Our findings indicate that no microphase separation of the DES components is induced by confinement, and that the primary dynamical motions observed in the bulk state persist, albeit with a marked slowdown in local molecular dynamics. These insights have direct implications for the diverse applications of deep eutectic solvents (DES), particularly in gas absorption, extraction processes and the development of energy-related nanomaterials.

## MATERIALS & METHODS

Fully hydrogenated anhydrous ethylene glycol (EG) ($C_2O_2H_6$, 99.8% purity) and choline chloride (ChCl) ($C_5H_{14}NO$ >99% purity) were purchased from Sigma-Aldrich. Partially deuterated ethylene glycol (1.1.2.2. $D_4$, 99% D) and choline chloride (trimethyl D9, 98% D) were obtained from Eurisotop. The chemicals were directly used to prepare DES without additional purification. Ethaline mixtures were prepared by combining EG and ChCl in specific molar ratios. For quasi-elastic neutron scattering (QENS), we prepared two mixtures with a specific molar ratio 1:4 (i.e., one mole of ChCl combined with 4 moles of EG) but employing different isotopic compositions so that one of the two constituents is deuterated and the other is fully hydrogenated. They are later denoted ChCl(H)/EG($D_4$) for hydrogenated choline chloride combined with deuterated ethylene glycol and ChCl($D_9$)/EG(H) for deuterated choline chloride combined with hydrogenated ethylene glycol. For small-angle neutron scattering (SANS), a different molar ratio 1:2.3 was selected in order to achieve equal volume fraction between ChCl and EG ($X_{volume} = 0.5$) and thus, maximize the sensitivity of the SANS method to any hypothetical microphase separation. The mixtures were prepared in a glove box and mechanically agitated at ~60 °C for 30 minutes until a clear, homogeneous liquid phase was obtained.

Deuteration plays a crucial role in the neutron scattering methods. First, it is noteworthy that the coherent scattering length of hydrogen and deuterium atoms have different signs ($^1$H, $b_{coh}$ = -3.74 fm) and ($^2$H, $b_{coh}$ = 6.67 fm). As a result, mixtures with different isotopic composition have different scattering length densities. We leveraged this phenomenon to modulate the scattering contrast in our small-angle neutron scattering (SANS) experiments. Second, Hydrogen ($^1$H) exhibits a very large incoherent scattering cross section ($\sigma_{inc}$= 80.27 barns), which can dominate the total signal. By substituting hydrogen with deuterium ($^2$H, $\sigma_{inc}$= 2.05 barns), the incoherent contribution of the labeled molecule is drastically reduced, enhancing the relative visibility of the hydrogen-rich counterpart. This phenomenon is particularly useful in our quasielastic neutron scattering experiments (QENS), in order to selectively tune the incoherent scattering contribution of each component. The coherent and incoherent scattering cross sections of the isotopically labeled molecules used here are provided in Table S1.[31] Accordingly, in ChCl(H)/EG($D_4$), the neutron cross section is dominated by the cholinium cation, with ethylene glycol contributing roughly one third of the total. Conversely, in ChCl($D_9$)/EG(H), the contributions are inverted, with ~75% of the incoherent scattering arising from EG.

SBA-15 mesoporous silicas were synthesized following a procedure reported previously.[40] A nonionic triblock copolymer (Pluronic P123, $EO_{20}$–$PO_{70}$–$EO_{20}$) acted as a structure-directing agent, yielding a hexagonally ordered mesostructured with a pore diameter of 8.1 nm. After calcination, the matrices were dried under primary vacuum at 120 °C for 12 hours. For MCM-41, hexadecyl ammonium bromide templated a similar hexagonal arrangement of channels with a smaller pore diameter of 3.5 nm. Pores radii were verified by nitrogen adsorption (KJS)[41] and neutron diffraction. Approximately 220 mg of powder was packed into glass tubes and filled via liquid imbibition with the DES solutions. The systems were equilibrated for a minimum of 3 days, before being transferred to the neutron sample cells.

Neutron diffraction experiments were carried out on the high-resolution D16 diffractometer at the Institut Laue-Langevin (ILL, Grenoble, France). We used a sample–detector distance of 1 m and an incident neutron wavelength of 4.5 Å, which allowed us to measure momentum transfers from 0.03 to 1.9 $Å^{-1}$. Empty pores materials were placed in cylindrical aluminum cells (5 mm inner diameter). The cells were then filled by imbibition with known masses of ethaline DES, chosen to fully saturate the pore volume. Finally, the diffraction data were corrected for detector efficiency, the empty-cell background was subtracted, and the spectra were normalized.

Quasielastic neutron scattering (QENS) measurements were carried out at the Institut Laue-Langevin (ILL, Grenoble, France) using two complementary spectrometers with distinct energy resolutions.[42] For both experiments, we used the same samples filled in rectangular flat aluminum cells. The disk-chopper time-of-flight spectrometer IN5B was employed with an incident wavelength of 4.9 Å. Under these conditions, the energy resolution at the elastic line was $\delta E \approx 80$ μeV (FWHM), corresponding to a characteristic timescale of ~10 ps ($t = \hbar/\delta E$). The quasielastic signal retained for analysis spanned an energy window of $-1$ meV $\leq E \leq 1$ meV and a momentum transfer range of $0.2 \leq Q \leq 2.1$ Å$^{-1}$. To access slower dynamics, the IN16B backscattering spectrometer was used, offering an energy resolution approximately two orders of magnitude higher. This configuration employed unpolished Si(111) monochromators and analyzers to select an incident wavelength of 6.27 Å, yielding $\delta E \approx 0.75$ μeV and thus probing a timescale on the order of 1 ns. IN16B was operated in both elastic and inelastic fixed-window scan (FWS) modes at energy transfers of 0, 3, and 6 μeV.

The combined use of time-of-flight and backscattering techniques enabled exploration of ethaline dynamics across a broad temporal window, from ~10 ps to ~1 ns. Sample temperature was controlled with a cryofurnace. For IN5B, measurements were conducted after thermal equilibration at 300 K, 325 K, and 350 K, whereas IN16B experiments employed a continuous cooling ramp from 350 K down to 10 K with a cooling rate of about 0.7 K/min. Both isotopically substituted DES mixtures confined in SBA-15 and MCM-41 were investigated under identical conditions.

Raw spectra were normalized to the incident neutron flux and corrected for detector efficiency using vanadium as a standard. Background contributions from the spectrometer and empty sample holders were subtracted prior to conversion of the measured intensity into the dynamic structure factor $S(Q,\omega)$. Data treatment and fixed-window analysis were performed with the MANTID software package,[43] while frequency-domain fitting of $S(Q,\omega)$ was carried out using the QENSH program developed at the Laboratoire Léon Brillouin (LLB, Saclay, France).

# RESULTS & DISCUSSIONS

## Structural profile: Small angle neutron scattering

In this section, we investigate whether choline chloride/ethylene glycol (ChCl/EG) ethaline DES mixtures confined within SBA-15 and MCM-41 mesoporous silica adopt homogeneous distributions by modeling the small angle neutron scattering intensity.

The diffraction patterns of SBA-15 and MCM-41 silica matrices, either empty or filled with isotopic ethaline mixtures are displayed in Fig. 1. They revealed distinct Bragg peaks, characteristic of the crystalline arrangement of the pores. These peaks were indexed consistent with the well-established hexagonal configuration of parallel nanochannels. The constant $q$-positions of the empty and filled pores Bragg reflections indicate the preservation of the hexagonal mesostructured of the silica mesoporous upon filling. [40,44,45] The strains due to capillary/adsorption stresses were not observed as they are far below the resolution of our SANS setup.

The most noticeable effect of pore filling was a reduction of the Bragg peak intensities, which reflects the change in scattering contrast. Under the conditions of a homogeneous liquid composition within the pores, the intensity of the different Bragg peaks of the filled samples $I^{Homo}(q)$ should simply be scaled with respect to the Bragg peak intensities of the empty hosts $I^{empty}(q)$ by the difference in scattering length density (SLD) $\Delta\rho$ between the filled pore volume and the pore wall (Eq. 1). The pore wall is composed of amorphous silica with a uniform density so that the SLD of silica is $\rho_{SiO_2} = 34.64 \times 10^9 cm^{-2}$.

$$I^{Homo}(q) = \left(\frac{\Delta\rho}{\rho_{SiO_2}}\right)^2 I^{empty}(q) \tag{1}$$

The scattered intensity of the filled SBA-15 presented a dependence on the isotopic composition of the two ethaline DES mixtures [Fig. 1(a)]. A straightforward origin of these changes may be the slightly different SLD of the two systems ($\rho_{\mathrm{ChCl(D9)/EG(H)}} = 25.35 \times 10^{9} cm^{-2}$, $\rho_{\mathrm{ChCl(H)/EG(D4)}} = 28.32 \times 10^{9} cm^{-2}$). Unlike SBA-15, MCM-41 matrices filled with ethaline exhibited a significantly smaller reduction in Bragg reflection intensities [Fig. 1(b)], which is inconsistent with a simple interpretation in terms of scattering contrast.

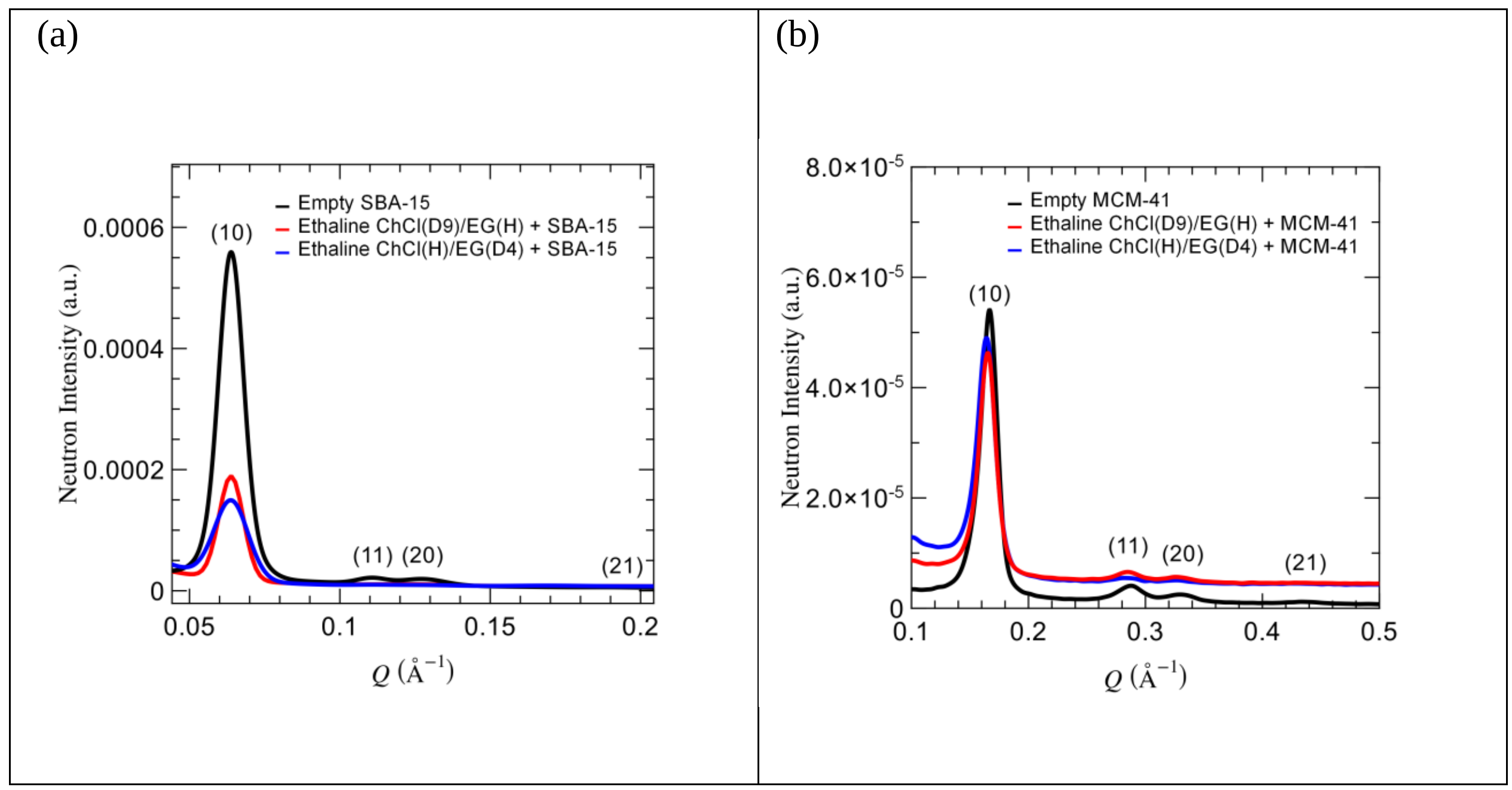

**FIG. 1**. Neutron diffraction patterns of (a) SBA-15 and (b) MCM-41. Empty matrix (black solid line), matrix filled with ethaline ChCl($D_9$)/EG(H) (red solid line), and matrix filled with ethaline ChCl(H)/EG($D_4$) (blue solid line).

These findings highlight the need for quantitative modeling of the scattered intensity in order to assess the extent of pore filling and to elucidate the nanostructure adopted by the deep eutectic solvent embedded within the porous volume. More precisely, beyond determining the filling fraction, the key question is whether the confined DES mixtures remain homogeneous throughout the pore volume or instead develop microphase-separated domains.

The methodological framework employed here to model the scattered intensity of the confined systems was originally developed for confined toluene/tert-butanol binary mixtures[21,22] and glycerol aqueous solutions,[46] and is recalled in the Supplementary Information.

The neutron diffraction intensity emanating from the silica materials, conceptualized as a periodic triangular array of cylindrical pores, can be expressed as follows:

$$I(q) = K\, S(q)\, |F(q)|^2 \tag{2}$$

where $K$ is an instrumental normalization factor, $S(q)$ denotes the powder-averaged structure factor of the hexagonal lattice, and $F(q)$ represents the scattering amplitude associated with the form factor of a single pore.

In a first step, the experimental integrated intensity $\tilde{I}(q_{hk})$ of each Bragg reflection was determined using an analytical fitting function that reproduces the experimental diffractograms with excellent accuracy. In a second step, the integrated intensities $\tilde{I}(q_{hk})$ were compared to the predicted intensities derived from various models, for which the form factor $|F(q)|^2$ can be calculated analytically (see Supporting Information).

For a given porous matrix, both $K$ and $S(q)$ are fixed quantities and were determined independently from the diffraction patterns of the corresponding empty matrices. For both dry matrices, the experimentally measured Bragg reflections closely matched the model form factor of an empty cylindrical pore.[46] Best fits were obtained for pore radii $R_{pore} = 4.05$ nm (for dry SBA-15) and $R_{pore} = 1.75$ nm (for dry MCM-41). The Debye–Waller factors accounting for static structural disorder were also fitted simultaneously. The values of the static mean square displacements of the cylinders from their regular lattice positions were $\langle u^2 \rangle = 0.5$ Å$^2$ and $\langle u^2 \rangle = 2$ Å$^2$, respectively.[46]

The only term that varies upon pore filling is therefore the form factor $F(q)$, which governs the modulation of the Bragg peak intensities. Its evolution reflects the spatial organization of the confined phase as well as its chemical composition and isotopic contrast. Two limiting radial concentration profiles were considered. The first corresponds to a uniform distribution of the isotopic species within the pore volume, i.e., a homogeneous filling characterized by a single average composition across the entire cross-section. In this case, the diffraction pattern of the filled matrix differs from that of the empty framework only through a modification of the neutron scattering contrast. The second scenario assumes partial microphase segregation inside the pores, resulting in a cylindrical core–shell geometry. In this configuration, one component is preferentially adsorbed at the pore surface, forming an interfacial shell, while the other predominantly occupies the central region. Analytical expressions for both models are provided in the Supplementary Information.

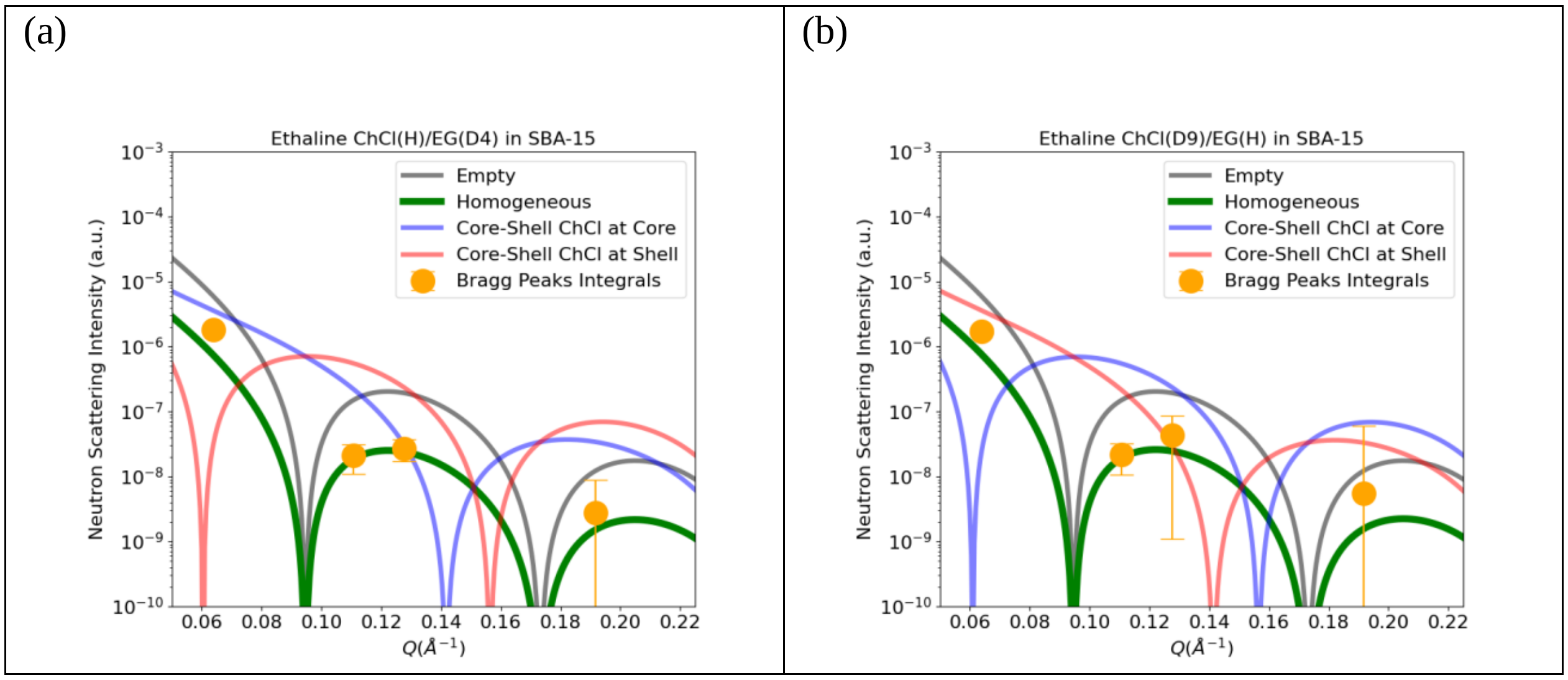

**FIG. 2.** Experimental integrated intensities of four Bragg peaks of the SBA-15 filled with ethaline with different isotopic mixtures (a) ethaline ChCl(H)/EG($D_4$), (b) ethaline ChCl($D_9$)/EG(H). The predicted forms factors are illustrated for empty SBA-15 (grey solid line),

homogeneous model (green solid line), core-shell model with ChCl at shell (red solid line) and ChCl at core (blue solid line).

In Fig 2, the different model predictions are compared with the experimentally observed Bragg peak intensities for both mixtures, ChCl($D_9$)/EG(H) and ChCl(H)/EG($D_4$) confined in SBA-15. The integrated Bragg intensities are best described by a model assuming a homogeneous radial distribution of the scattering length density of ethaline inside the pores. In contrast, models based on compositionally distinct shell or core domains fail to align with the experimental observations. This suggests that both choline chloride and ethylene glycol are uniformly distributed across the pore cross-section, rather than one component being selectively adsorbed at the silica surface. In other words, the DES mixture remains structurally homogeneous radially, with no evidence of a two-phase microphase separation.

Similar to SBA-15, filling the MCM-41 pores resulted in a *Q*-independent reduction in the intensity of its Bragg reflections. This observation excludes models based on a core-shell radial distribution of the two ethaline components. However, the smaller reduction in intensity observed after filling MCM-41, compared to SBA-15, could not be quantitatively reproduced with the homogeneous model. This indicates that only a fraction of the pores was completely filled by the DES mixture. This hypothesis is actually confirmed by the good agreement obtained with a radially homogeneous model that assumes that only 60% of the total porosity was actually filled [Fig. S3]. The partially unfilled porosity in MCM-41 may suggest exceptionally slow imbibition kinetics of ethaline within the smaller pores, likely due to the high viscosity and low vapor pressure of DES.

The observation that the composition of the ethaline mixture remains homogeneous across the pore diameter is a significant difference from other binary mixtures, in particular tert-butanol/toluene, which exhibit microphase separation when confined in MCM-41 and SBA-14.[21,23] For the latter mixture, the formation of a core-shell organization in nanopores was attributed to the dissimilar interactions between the two components and the silica pore wall.[47] Tert-butanol being a polar H-bonded liquid, it forms preferential interactions with the hydrophilic surface of silica, which drives segregation of the weakly polar van der Waals toluene molecules into the pore center. A distinct behavior was observed in binary systems composed of two polar, hydrogen-bonding components, such as the water–glycerol mixture.[46] In this case, the system constituents remain homogeneously distributed within the confined liquid state. This can be interpreted as the cumulative result of strong water–glycerol interactions, which preserve the cohesive associations between the two components, combined with the similar affinity of both molecules for the hydrophilic pore surface. In water–glycerol mixtures, the formation of core-shell structure was observed only at low temperature upon ice formation. Unlike the tert-butanol/toluene system, where micro-segregation is driven by intermolecular and interfacial interactions, this phenomenon arises from thermodynamic phase separation induced by ice crystallization.

In contrast, ionic liquids (ILs) typically adopt distinct layered structures when confined in porous geometries.[26] This specific structural organization arises from the interplay of various interactions, with electrostatic forces playing a dominant role. This corresponds to a periodic, alternating density profile of cation–anion pairs near solid surfaces, a phenomenon not limited to porous materials, but also observed at interfaces with flat solids.[48]

Two main reasons can be invoked to interpret the observation of a uniform distribution of the DES component across the pore. First, unlike a TBA/toluene-like pair (where one molecule is strongly attracted to the surface and the other is not), both constituents of ethaline are polar/hydrogen-bonding species that both present favorable interactions with the hydrophilic silica surface. Choline cations and chloride anions likely adsorb via electrostatic or H-bond interactions, and ethylene glycol is a highly polar hydrogen-bond donor that also readily H-bonds to silanol groups. Because both components have affinity for the surface, none of them is exclusively preventing the development of a core-shell composition gradient. From this point of view, the situation encountered for ethaline shares similarity with the case of water-glycerol mentioned above.[46] Secondly, neutron scattering studies have reported complexed ionic clusters for choline chloride-based DESs with different HBDs.[49–51] Specifically, ethaline forms an interpenetrated hydrogen-bond network dominated by EG–Cl interactions, without evidence of large-scale domain segregation.[52] Only when ethylene glycol is present in excess of the eutectic ratio (e.g., 1:4 compositions) or when water partially replaces EG in the solvation shell does a weak low-$Q$ feature emerge in SANS data, attributable to mesoscopic segregation of EG. We infer that the DES behaves as a highly cohesive phase in the bulk with structure governed by strong EG-ChCl interactions, which prevent any microscopic demixing under confinement. While our low-$Q$ neutron scattering results confirm a homogeneous composition and rule out radial phase separation, they do not provide insight into the specific nature of short-range order between neighboring molecules within the first coordination shell. Therefore, we cannot dismiss the potential presence of weak interfacial layering, as has been observed in ionic liquids and confined hydrogen-bonded systems.[26,53,54]

## Local picosecond timescale molecular dynamics

The QENS spectra of ethaline (1:4) were measured by neutron quasielastic scattering on the time-of-flight instrument IN5B for the two distinct isotopic compositions and confining matrices. The spectra are compared with those reported for the same DES systems in the bulk state that were acquired in the same experimental conditions by Kamar et al.[31] [Figs. 3 (a) and 3(b)]. The spectra of the DES mixtures confined in SBA-15 are presented in Figs. 3(c) and 3(d) for the three investigated temperatures at a representative momentum transfer of $Q$ = 1.5 Å$^{-1}$. For MCM-41 filled matrices, the equivalent spectra are illustrated in Figs. 3(e) and 3(f). On a qualitative level, the spectra of the confined systems present the same features as their bulk counterparts. We observe a quasielastic broadening extending over [-1:1] meV, that reflects the molecular motions on a typical timescale of 1-100 picoseconds. Because the incoherent scattering cross section of hydrogen dominates the signal (see Table S1), the observed broadening is attributed to the self-part of the dynamic structure factor $S(Q,\omega)$.[55] A progressive narrowing of the quasielastic peak was observed upon cooling from 350 K to 298 K, which is consistent with a gradual slowing of the liquid molecular dynamics with decreasing temperature. The quasielastic lineshapes observed in the confined systems [Fig. 3(c, d)] are narrower compared to those of the bulk systems [Fig. 3(a, b)], pointing to slower dynamical processes under nanoconfinement. Similar observations were made for the ethaline mixtures confined in MCM-41 [Fig. 3(e, f)].

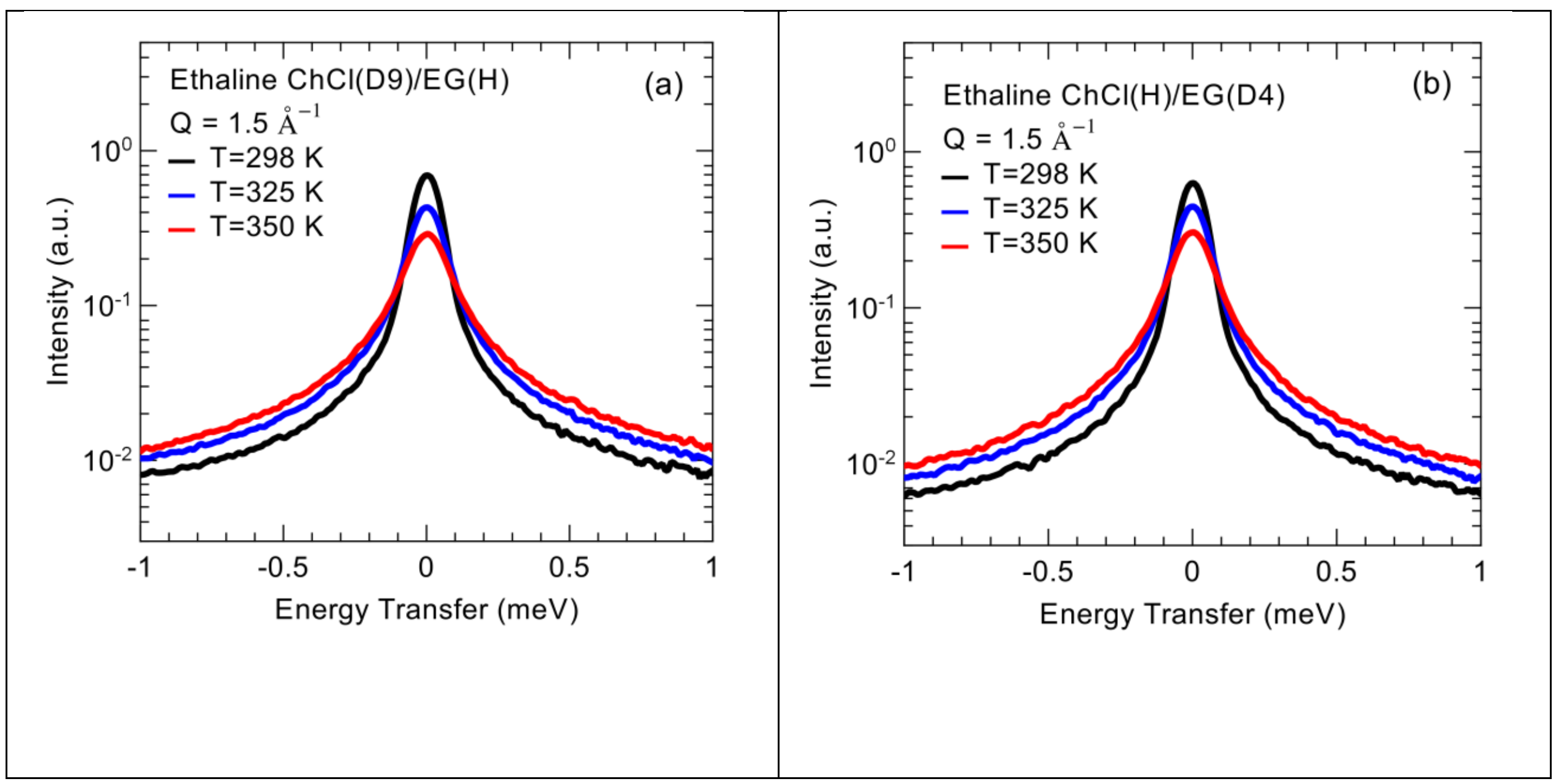

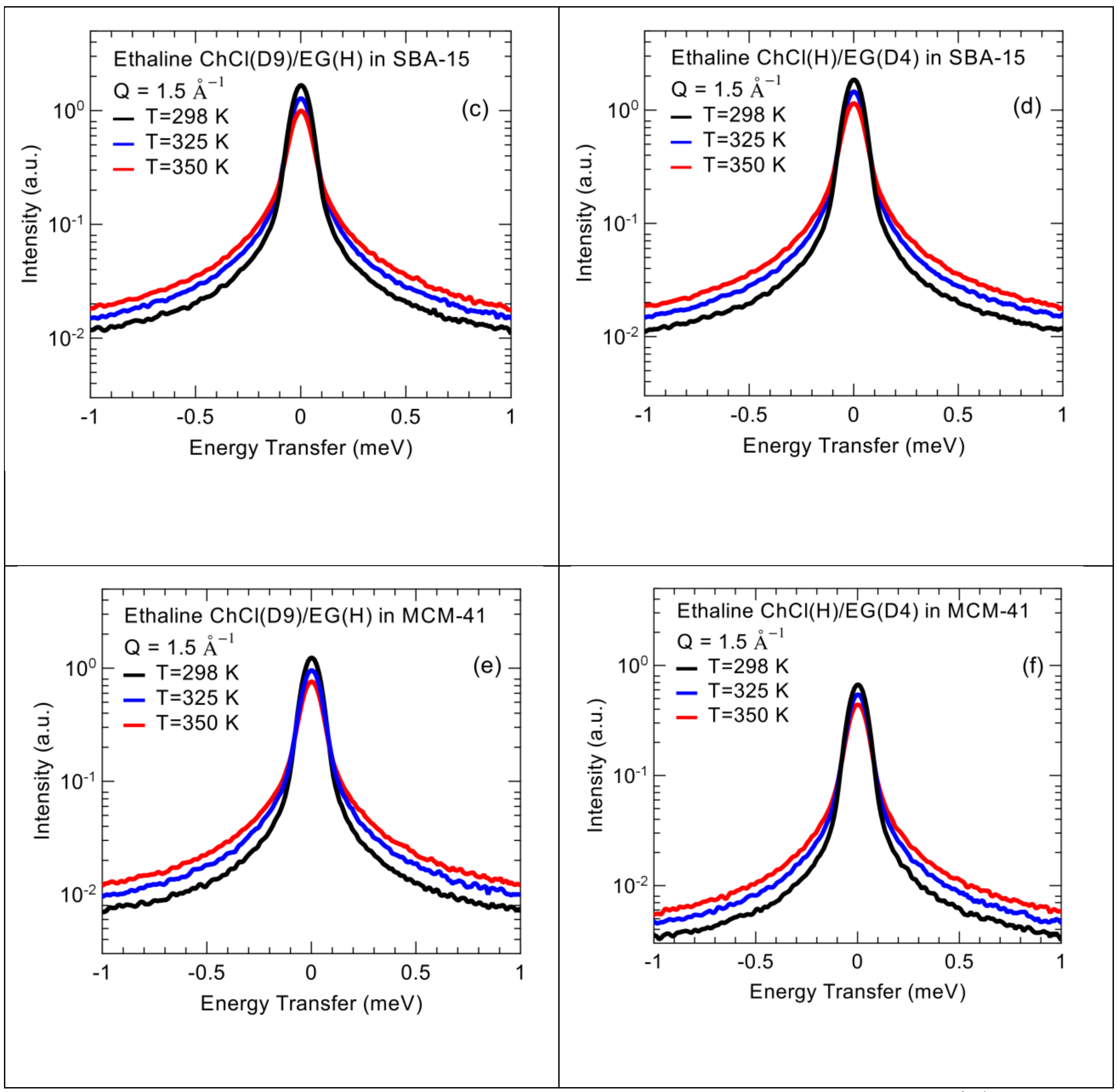

**FIG. 3.** Temperature dependence of the QENS spectra measured on IN5B at $Q = 1.5$ Å$^{-1}$ of ethaline ChCl($D_9$)/EG(H) and ChCl(H)/EG($D_4$), (a, b) in the bulk, (c, d) confined in SBA-15, (e, f) confined in MCM-41. Bulk Ethaline results are reproduced from The Journal of Chemical Physics **163**(13), 134506 (2025), with the permission of AIP Publishing.

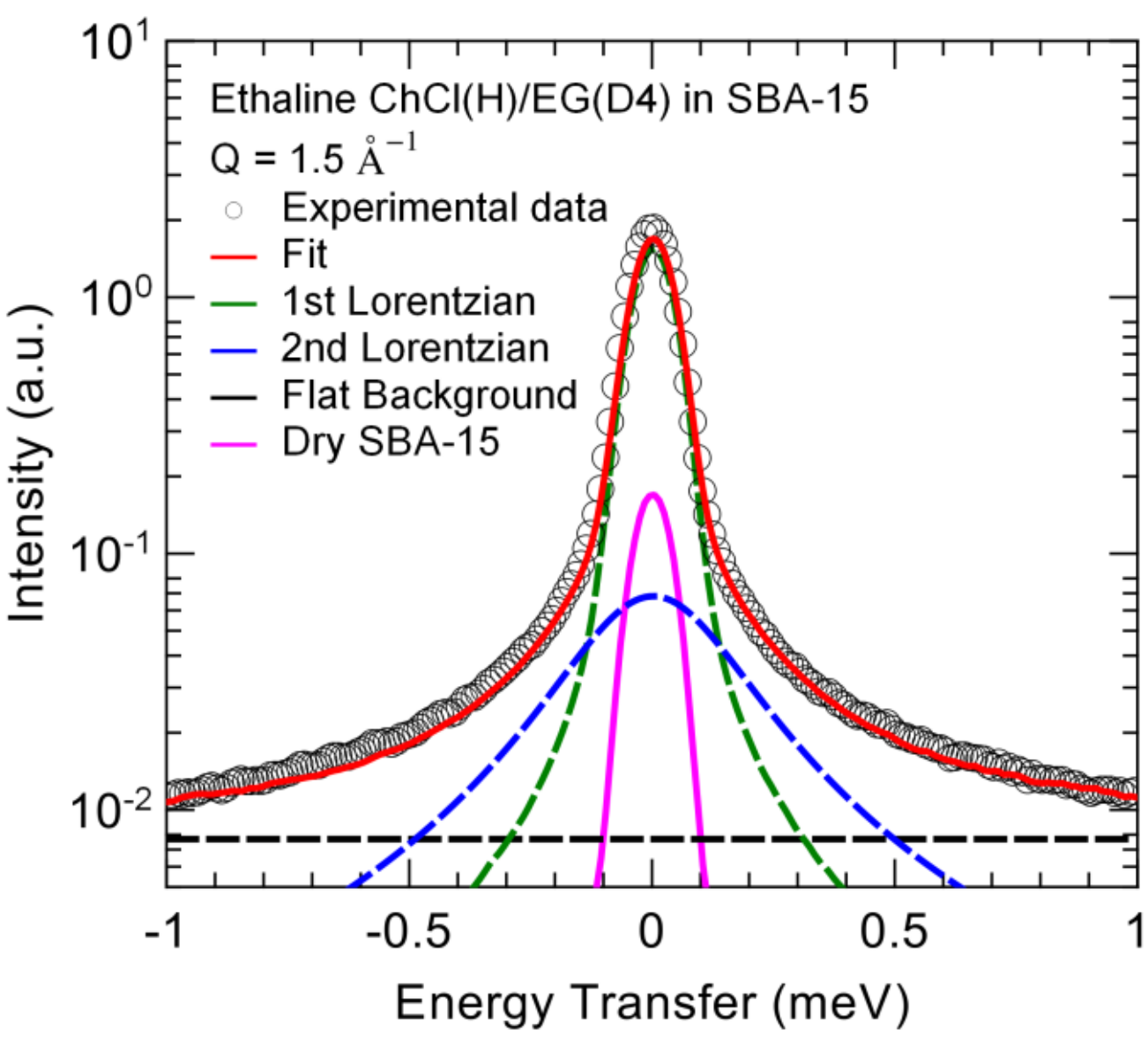

**FIG. 4.** Experimental spectrum of ethaline ChCl(H)/EG($D_4$) confined in SBA-15 at $T$ = 298 K and at $Q$ = 1.5Å$^{-1}$ (open circles). Fit of a model (solid red line) comprising a sharp Lorentzian (green dashed line), second broad Lorentzian (blue dashed line) and flat background (black dashed line). Experimental spectrum of the empty SBA-15 (solid pink line).

For bulk ethaline, it was shown that the dynamic structure factor measured on this energy range arose from two distinct motions:[31] A localized fast motion, modeled as $S_L(Q,\omega) = A_L(Q)\delta(\omega) + (1 - A_L(Q))\, L_L(Q,\omega)$ and, a translational component modeled as $S_T(Q,\omega) = L_T(Q,\omega)$, where $L_X(Q,\omega)$ denotes a Lorentzian with half width at half maximum (HWHM) $\Gamma_X$. This decomposition is widely accepted and has been reported in QENS studies of liquids, including deep eutectic solvents (DESs), such as choline chloride-based DES and alkylamide–lithium perchlorate systems.[31,32,34–37] The localized motions encompass rattling within a transient cage, rotations/librations, conformational changes, and cage relaxation. Given the qualitative diversity and dynamic heterogeneity of these processes, a distribution of characteristic times and lengths is expected. Due to this complexity, individual microscopic parameters cannot be uniquely extracted from QENS data. Instead, simplified models with a reduced number of effective parameters are typically employed. In some studies, local motions were approximated as a single atomic translational diffusion process confined in a hard sphere with all hydrogen atoms assumed to have identical dynamics (same diffusion coefficient), while the heterogeneity is accounted for via a distribution of confining radii.[34–37] For ethaline, localized motion seems not well-described by confined Fickian diffusion on the studied time and length scales, as the observed that the quasielastic broadening ($\Gamma_L$) remains essentially constant across the accessible $Q$-range.[31] Instead, the dynamics were better characterized by a mean time $\tau_L = \frac{\hbar}{\Gamma_L}$, rather than effective local diffusion coefficient, suggesting they were dominated by molecular jumps and internal configurational degrees of freedom.

To ensure methodological consistency and a scientifically rigorous comparison of confinement effects, we applied the same methodology for confined Ethaline as previously validated for the bulk systems by Kamar et al.[31]. This approach actually yields excellent fits to the data, as illustrated for a specific $Q$-value in Fig. 4, and allows for a quantitative assessment of confinement effects.

In the following, we focus on the local fast dynamics, while the slower translational diffusion process is discussed in the next section, along with neutron backscattering results. The dynamic structure factor of the local motion, $S_L(Q,\omega)$, contains two important pieces of information: - the relaxation time $\left(\tau_L = \frac{\hbar}{\Gamma_L}\right)$, which characterizes the timescale of local dynamics; - the elastic incoherent structure factor (EISF), $A_L(Q)$, which provides insight into the spatial trajectory of the motion.

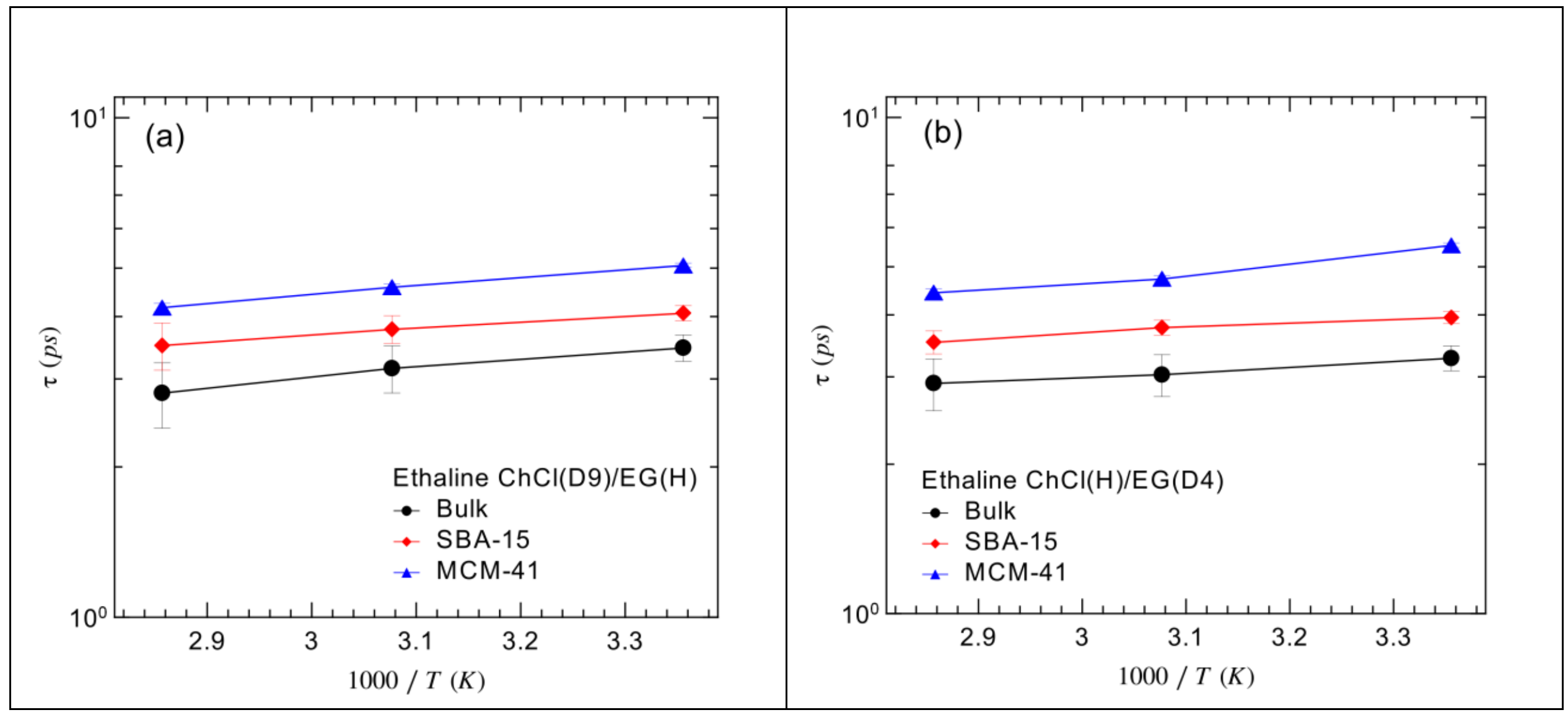

**FIG. 5.** Relaxation time ($\tau_L$) of the local motion for (a) ethaline ChCl($D_9$)/EG(H) and (b) ethaline ChCl(H)/EG($D_4$) bulk (black circles), confined in SBA-15 (red squares) and confined in MCM-41 (blue triangles).

The width $\Gamma_L$ of the broad Lorentzian $L_L(Q, \omega)$ did not present any significant $Q$-dependence, as shown in Fig. S10, which is consistent with the non-dispersive nature of the localized motion. The obtained values of $\tau_L$are illustrated in Fig. 5. In bulk ethaline, the relaxation time ranges between 2.5 and 3.5 ps, showing only a slight Arrhenius temperature dependence. In confined ethaline, we observe an increase in the relaxation time of approximately 20% for SBA-15 ($\tau_L \approx$ 3.5-4 ps) and up to 50% for MCM-41 ($\tau_L \approx$ 4-6 ps). The amplitude of these changes is comparable for both isotopic compositions. This demonstrates that the two components of ethaline, ChCl and EG, experience a pronounced slowdown of their picosecond-scale local motions that intensifies as the pore size decreases.

The trajectory associated to this fast motion can be characterized by the EISFs illustrated in Fig. 6 for the ethaline in the bulk state and confined in the two matrices. We analyzed the EISF of the confined mixtures using the model proposed by Kamar et al. for bulk ethaline,[31] which is defined as follows:

$$A_L(Q) = \exp\left(\frac{-\langle u^2\rangle Q^2}{6}\right)[\,(1 - x_{OH} - x_{Me}) + x_{OH}A_{OH}(Q) + x_{Me}A_{Me}(Q)] \qquad (14)$$

where $\langle u^2\rangle$ is the mean-squared displacement associated with the Gaussian localized motion of the whole molecule; $x_{OH}$ and $x_{Me}$ are the molar fractions of hydroxyl and methyl protons; and $A_{OH}(Q)$ and $A_{Me}(Q)$ are the standard 2-site and 3-site jump EISFs, respectively. For mixtures with deuterated methyl protons the expression reduces by setting $x_{Me} = 0$. In this model, the Gaussian prefactor accounts for the librational and rattling motions of molecules confined within a cage formed by neighboring molecules, creating a transient harmonic potential. This can likewise be interpreted as molecular diffusion confined in a basin with soft boundaries.[55,56] However, as discussed earlier, the restricted motions in ethaline do not strictly follow the confined Fickian diffusion model, which predicts a distinct $Q$-dependence for the quasielastic broadening ($\Gamma_L$), and rather suggest librational and configurational molecular jumps. In the bulk state, the value of the mean displacement $R = \sqrt{\langle u^2\rangle}$ is typically 1 Å for ethaline.[31] This relatively small value supports that the attribution of the broad quasielastic component to local motions, with each DES ion/molecule being trapped in its cage formed by complexation and hydrogen bonding with neighboring components.[36] Notably, the radius of localized diffusion of hydrogen atoms within transient cages of an acetamide based DES made with lithium perchlorate was found in the range 0.9-1.7Å.[34] Additional degrees of freedom are assigned via $A_{OH}(Q)$ and $A_{Me}(Q)$ to conformational changes involving the methyl and hydroxyl groups. The onset of translational diffusion appears at longer times than these local processes, as indicated by the narrow quasielastic component. Indeed, it requires relaxation of the cage and molecule jumping from cage to cage. After a sufficiently large number a successive jumps, it eventually leads to Fickian diffusion.[36]

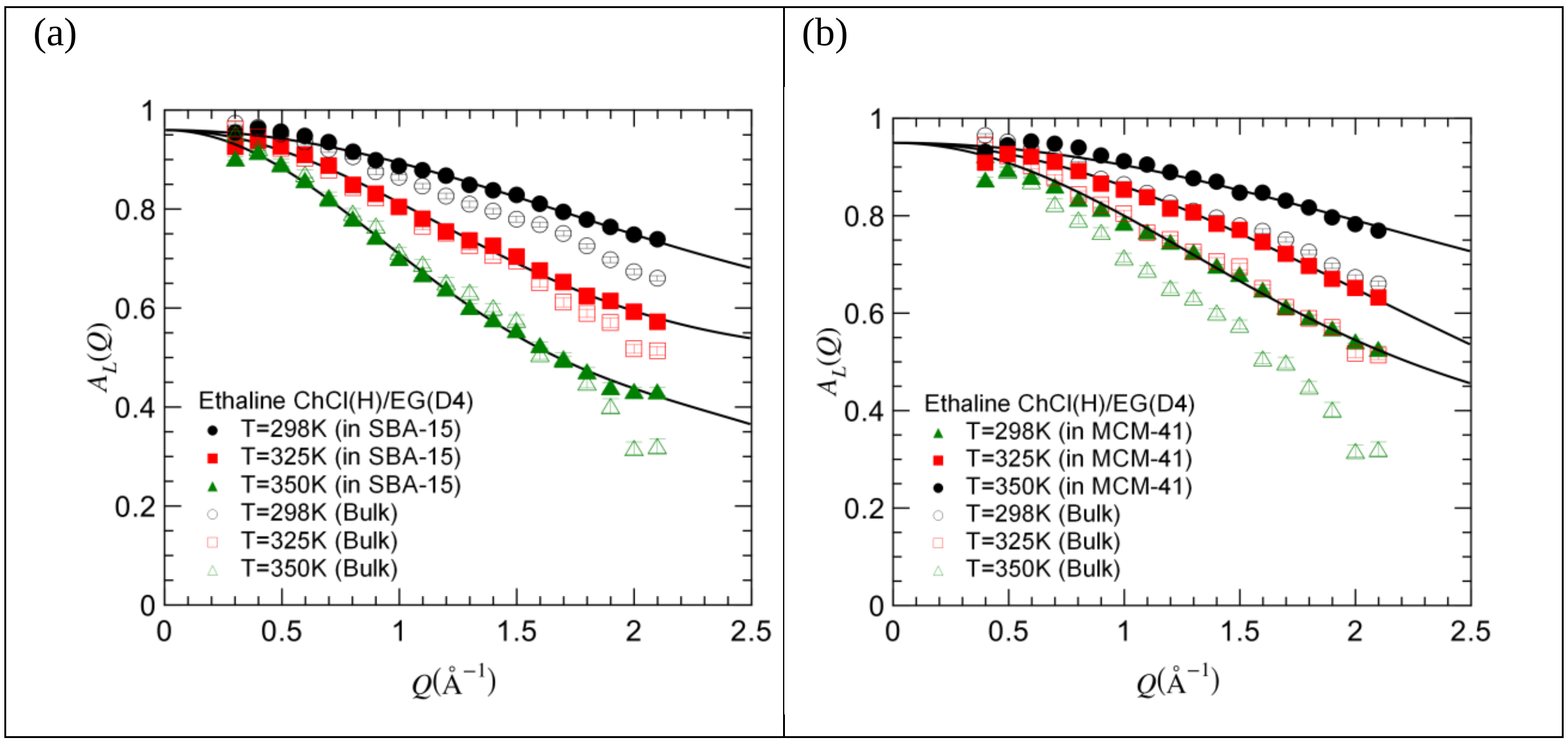

**FIG. 6.** Elastic incoherent structure factor $A_L(Q)$ of the local molecular motions at three different temperatures of ChCl(H)/EG(D4) confined in (a) SBA-15 and (b) MCM-41. In both panels, the confined mixture (filled symbols) and fit model (black solid lines) are compared to the bulk results (open symbols). Bulk Ethaline results are reproduced from The Journal of Chemical Physics **163**(13), 134506 (2025), with the permission of AIP Publishing.

The best fits obtained with this model are indicated as solid lines in Fig. 6 for the confined systems. In these fits, the intensity was rescaled to correct for a small (~5%) deviation of the experimental intensity from the theoretical expectation that the EISF should extrapolate to unity at $Q = 0$ Å$^{-1}$.[57] For all systems, we observe a systematic increase in intensity with decreasing the temperature from 350 K to 298 K. This indicates that the amplitude of the local motion decreases when cooling the system. Logically at low temperature, molecular motions are spatially more restricted in the local basin before cage-breaking enables translational diffusion. This interpretation is supported by a systematic decrease in the mean displacement $R = \sqrt{\langle u^2 \rangle}$, from approximately 1 Å at 350 K to fraction of an angstrom at 298 K [Table S2].

Compared to temperature, confinement has a less pronounced effect on the EISF values. This suggests that the trajectory of the local motions in the confined ethaline retains characteristics comparable to those observed in the bulk phase. This indicates that the molecules experience in the confinement a short-ranged local environment that is comparable to that of the bulk state. This is in accordance with findings from SANS experiments. Still, the intensity of the EISFs is somewhat larger in confinement compared to the bulk DES. This increase is modest for SBA-15, but its magnitude becomes more pronounced in MCM-41. Similar to temperature, this indicates that reducing the confining pore size limits the spatial amplitude of the local molecular dynamics.

## Translational diffusion

The translational diffusion of the DES mixtures has been analyzed combining the results obtained with the two spectrometers, IN5B and IN16B, that present complementary energy resolutions, in order to cover an extended temperature range. On IN5B, the translational dynamics were characterized by the sharp quasielastic component $S_T(Q, \omega) = L_T(Q, \omega)$, as defined in the precedent section. On IN16B, the translational dynamics were characterized by modelling the fixed-window scans.

The elastic and inelastic fixed-window scans (FWS) acquired on IN16B during cooling from 350 K to 10 K, summed over all momentum transfers $Q$, are shown in Fig. 7 for the DES mixtures confined in SBA-15. The scans of the DES mixtures confined in MCM-41 are illustrated in Fig. S7, while similar measurements were already presented for the bulk mixtures by Kamar et al.[31] The progressive increase of the elastic intensity with decreasing temperature reflects the gradual slowdown of molecular dynamics. Both isotopic compositions of DESs exhibited a smooth and continuous increase of elastic intensity, confirming the absence of crystallization and highlighting the glass-forming ability of ethaline, as confirmed by the absence of Bragg reflections [Fig S6]. This is consistent with findings from calorimetric and dielectric studies that measured the glass transition $Tg \approx 120$ K for ethaline confined in SBA-15 and $Tg \approx 118$ K for ethaline in MCM-41.[58,59]

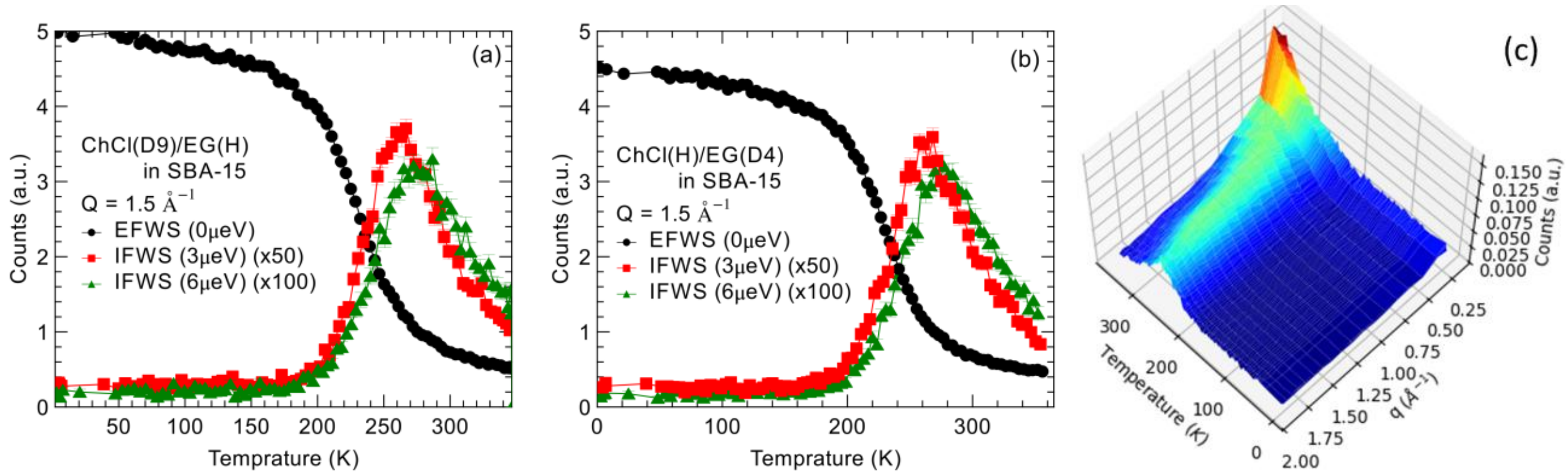

**FIG. 7.** Elastic and Inelastic Fixed Window Scans normalized to maximum elastic intensity measured on IN16B for (a) ethaline ChCl($D_9$)/EG(H) and, (b) ethaline ChCl(H)/EG($D_4$) confined in SBA-15 at $Q$ = 1.5 Å$^{-1}$. Intensity obtained at elastic position (black circle) and for two different energy off-sets 3 μeV (red square), and 6 μeV (green triangle). Scaling factors (50x and 100x) have been applied to IWFS for better clarity. (c) 3D plot of the IFWS of ethaline ChCl(H)/EG($D_4$) confined in SBA-15 as a function of temperature and momentum transfer.

The inelastic fixed-window scans (IFWS), measured at energy offsets of 3 μeV and 6 μeV, each revealed a single peak spanning temperature interval of approximately 240-330 K and 250-350 K, respectively, at half-maximum intensity as illustrated in Fig. 7. The peak maxima, centered around 270-280 K, shifted to higher temperatures with increasing energy offset and with decreasing $Q$. This behavior highlights the dispersive character of the underlying dynamics, attributed to the $Q$- dependent onset of translational diffusion processes within the dynamical window defined by the chosen energy offsets. By comparison, the broader quasielastic contribution detected on IN5B, associated with localized molecular motions, lies largely outside the accessible energy range of IN16B. Accordingly, a simplified model was employed to analyze the IFWS, expressed as:

$$I_{IFWS}^{fit}(Q,T) = A(Q)L_T(Q,\omega_{offset}) + BG(Q) \qquad (12)$$

where $L_T(Q,\omega_{offset})$ represents a Lorentzian term describing translational diffusion, and $BG(Q)$ accounts for the contribution from faster local motions. The IFWS intensities recorded at the two different offset energies $\hbar\omega_{offset}$ were fitted simultaneously using a single set of parameters, while each $Q$-value was treated independently. The linewidth of the Lorentzian was assumed to follow an Arrhenius dependence, $\Gamma_T(Q,T) = \Gamma_0(Q)exp\left(\frac{-E_a}{k_BT}\right)$ with a unique value of the activation energy $E_a$ for all $Q$ and offset values. Excellent fits of the data were obtained for all studied systems, as illustrated in supporting information.

The width $\Gamma_T$ of the narrow quasielastic component (HWHM) measured on IN5B at three temperatures (298 K, 325 K, and 350 K), plotted versus $Q^2$ for the two isotopically labeled ethaline mixtures in the bulk and confined in SBA-15 is illustrated in Fig. 8. The same illustrations are provided for the MCM-41 systems in Fig. S10. The values of $\Gamma_T$ at those temperatures were also inferred from the IFWS measured on IN16B by extrapolating the Arrhenius temperature dependence (see open symbols in Fig. 8). Although these target temperatures lie well above the IFWS peak region (≈250–285 K) and the resulting $\Gamma_T$ values (up to ~80 μeV) exceed the IN16B energy offsets by about an order of magnitude, the two instruments probing very different time scales remain in excellent mutual agreement.

In the bulk phase, Kamar et al.[31] reported that the linear variation of $\Gamma_T$ with $Q^2$ deviates at higher $Q$-values, approaching a constant plateau defined by $1/\tau_0$. They attributed this behavior to a breakdown of the continuous-diffusion assumption, as molecular displacements become comparable to intermolecular spacings, thereby favoring discrete jump motions over continuous dynamics. Accordingly, the linewidths were described with a standard jump-diffusion expression,

$$\Gamma_T(Q) = \frac{D_T Q^2}{1 + \tau_0 D_T Q^2} \tag{15}$$

where, $D_T$ is the translational diffusion coefficient and $\tau_0$ the mean residence time between jumps.

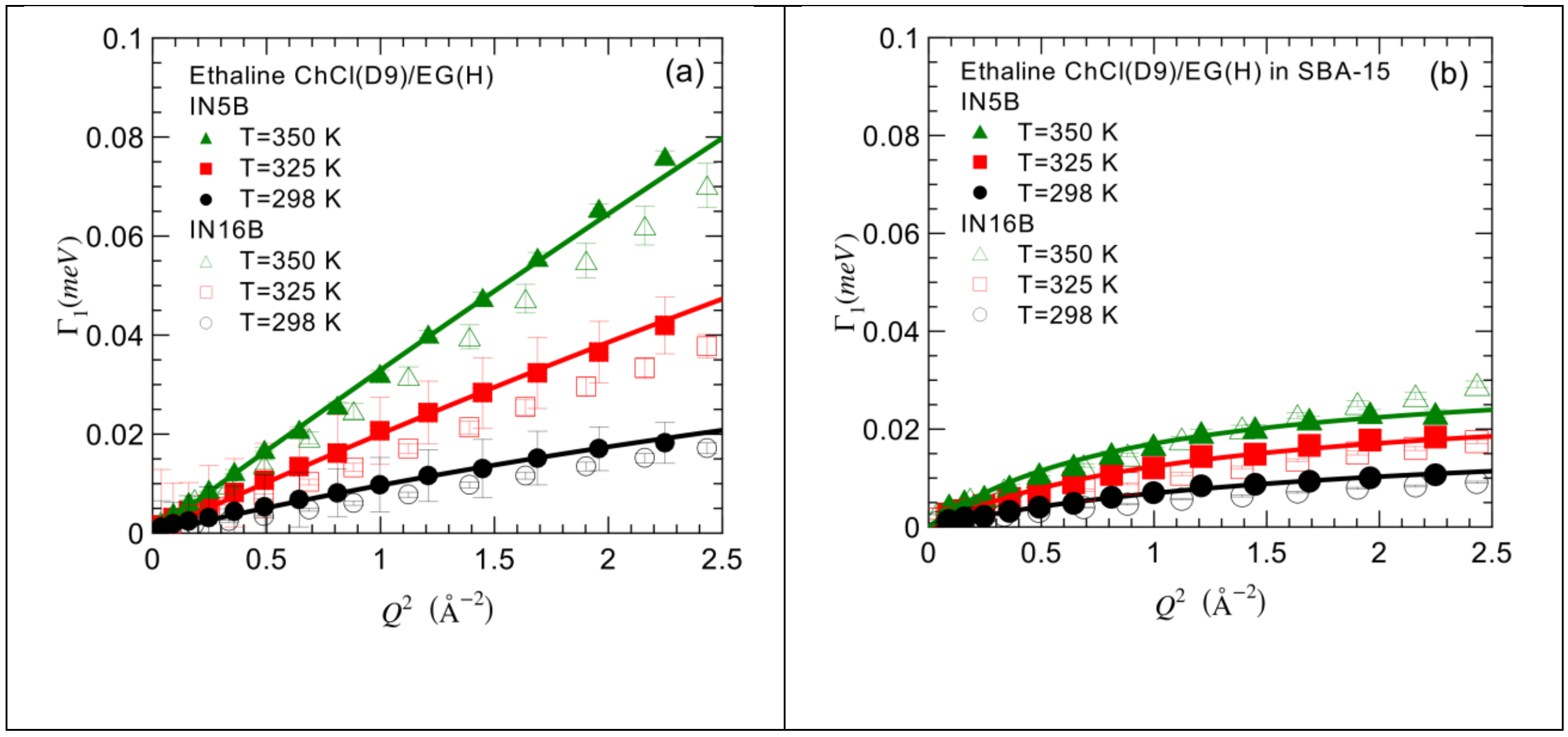

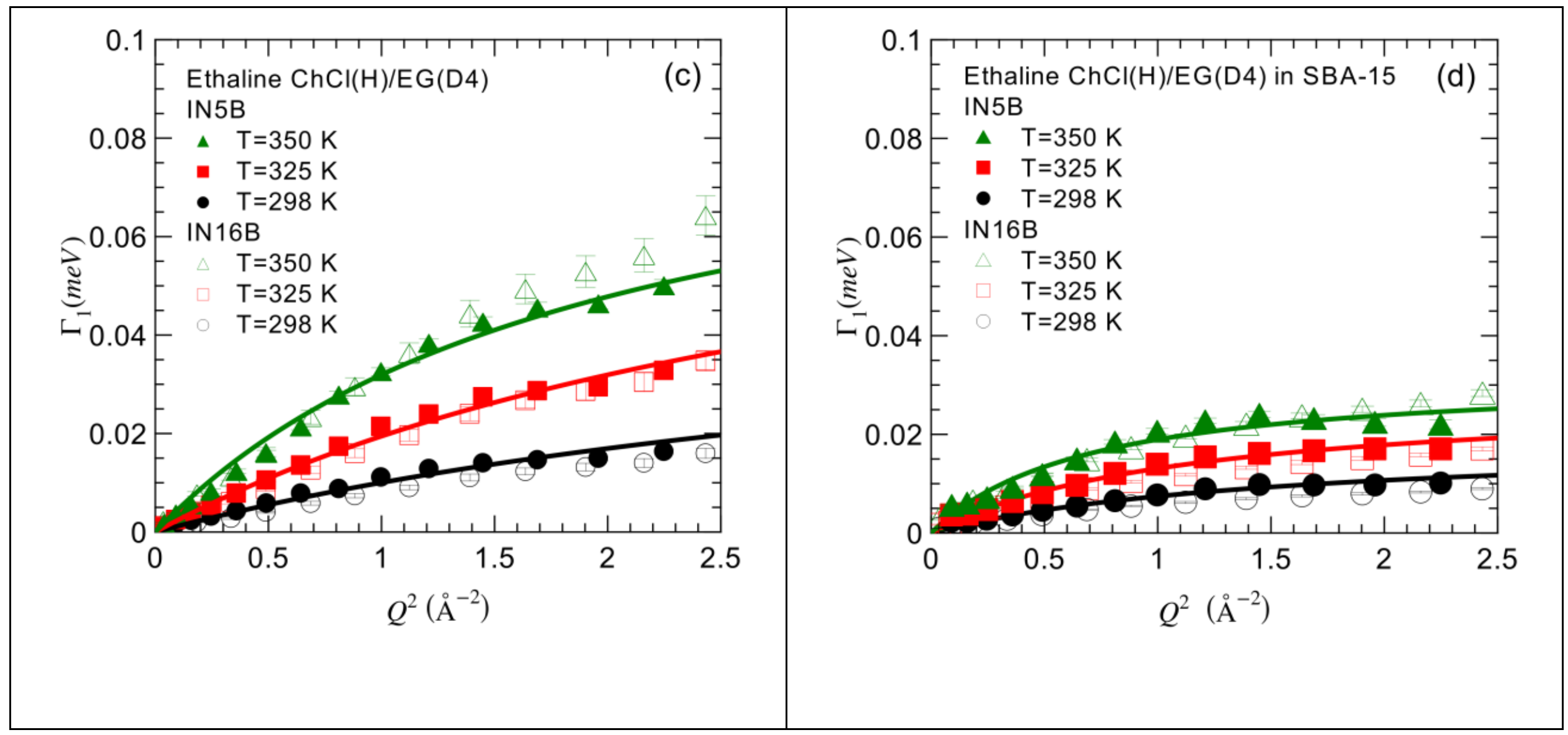

**FIG. 8.** Evolutions of the half width at half-maximum $\Gamma_T(Q,T)$ of the sharp Lorentzian attributed to translational diffusion as a function of $Q^2$ obtained from the fitting of QENS spectra measured on IN5B (filled symbols) and IN16B (open symbols) for (a) Ethaline ChCl(D9)/EG(H) and (c) Ethaline ChCl(H)/EG(D4) as bulk (b) and (d) in SBA-15 respectively. Fit using the jump diffusion model (thin solid lines) at three temperatures, from top to bottom 350 K, 325 K, and 298 K (solid lines). Bulk Ethaline results are reproduced from The Journal of Chemical Physics **163**(13), 134506 (2025), with the permission of AIP Publishing.

In the confined states, we observed qualitatively similar trends in the evolution of $\Gamma_T(Q)$. In fact, very good fits of the experimental data were obtained using the jump diffusion model (see solid lines in Fig. 8). However, at a quantitative level, a pronounced confinement effect emerges. This is particularly evident at high $Q$, where $\Gamma_T(Q)$ is markedly reduced in the confined binary mixture, signaling significantly slower dynamics. The fitted parameters (diffusion coefficient $D_T$, and residence time $\tau_0$) obtained for ethaline confined in SBA-15 are illustrated in Fig. 9 and compared to those of the bulk mixtures. A similar comparative analysis of the measurements for MCM-41 and SBA-15 is presented in Fig. S11, with all parameter values indicated in Tables S3-S5.

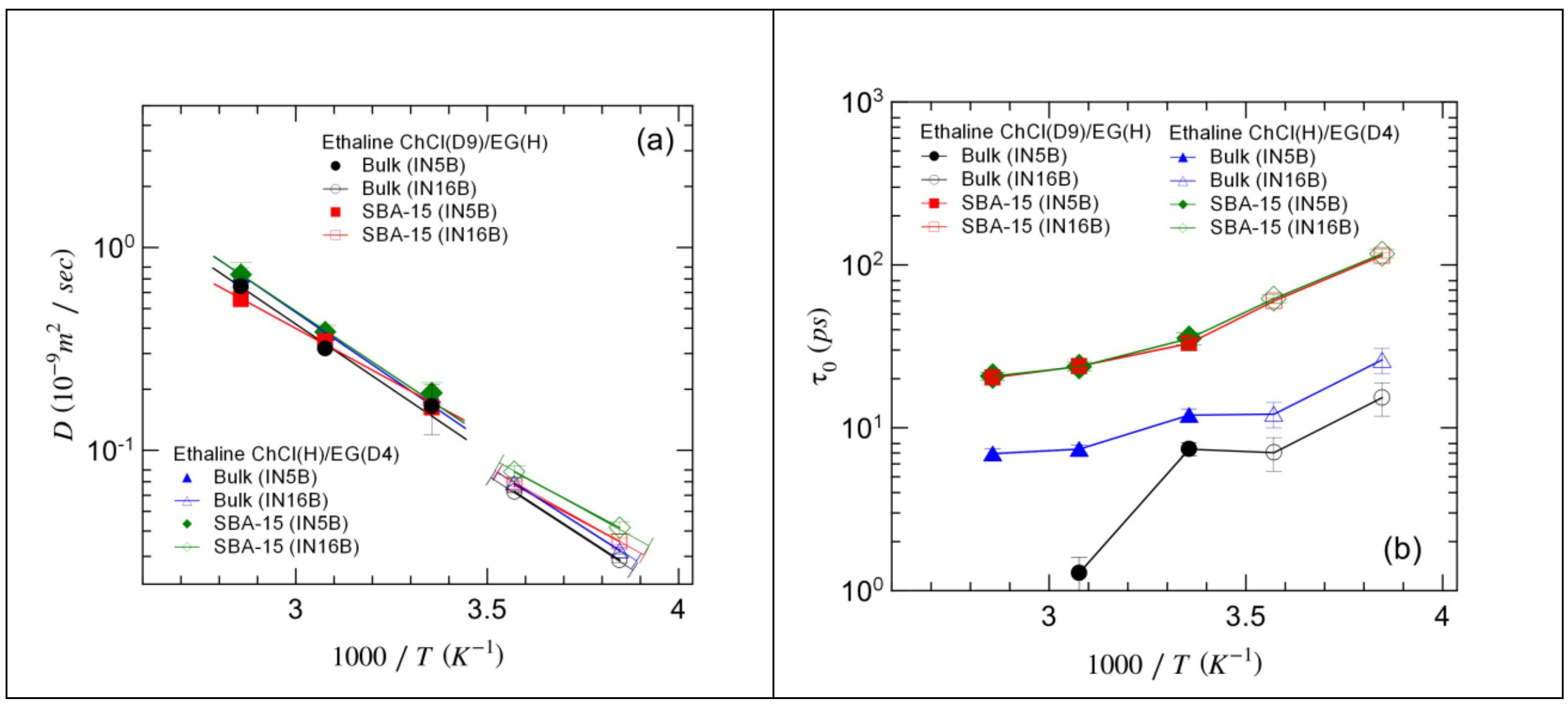

**FIG. 9.** (a) Translational diffusion coefficient $D_T$, (b) residence time $\tau_0$ of ethaline evaluated from the fit of IN5B spectra and IN16B IFWS. IN16B results are illustrated in the temperature range 250-285 K around the maxima of the IFWS (open symbols). Bulk Ethaline results are reproduced from The Journal of Chemical Physics **163**(13), 134506 (2025), with the permission of AIP Publishing.

It shows that nanoconfinement has only a marginal impact on the translational diffusion coefficient $D_T$ of ethaline. For SBA-15, the values $D_T$ are consistent with those of the bulk state, exhibiting non-systematic variations of less than 20%. Compared to SBA-15, the diffusivity of ethaline confined in MCM-41 appears systematically smaller, by about 10%. This observation might indicate a measurable slowdown effect of translational diffusion for the smallest pore size. If confirmed, this effect remains marginal and should be interpreted cautiously, as its magnitude is on the order of the typical experimental uncertainty.

In contrast to the diffusion coefficient, nanoconfinement significantly prolongs the residence time $\tau_0$ of ethaline molecules. Fig. 9 (b) reveals that the mean residence time increases by roughly a factor of 3-8 under confinement in SBA-15 relative to bulk. In MCM-41, a slightly more pronounced increase in the residence time $\tau_0$ is observed, in certain cases reaching up to a tenfold enhancement.

This increase of the residence time $\tau_0$ between two molecular jumps of the molecular center-of-mass echoes the increase in the local relaxation time $\tau_L$ (see the previous section) that is attributed to molecular motion inside the transient cage. It indicates that the environment formed by the neighboring molecules reduces the localized dynamics, and also forms a longer-lived cage in confinement compared to bulk. Similar confinement effects were observed for water confined in comparable porous silica.[60] In aqueous DES mixtures, the jump-diffusion mechanism has been associated with the formation of supramolecular complexes and the lifetime of H-bonds life time.[36] Given the observed increase in the residence time $\tau_0$ in ethaline, one may infer that confinement enhances the interactions between its constituents EG and ChCl, and, consequently, increases the stability of transient cages. Additionally, interfacial interactions mediated by H-bonds and electrostatic between the ethaline components and the hydrophilic surface of silica play a significant role on the jump-diffusion. This phenomenon aligns with previous observations of a considerable increase in the residence time $\tau_0$ for water confined in periodic mesoporous organosilicas with ionically charged surfaces.[61]

A different situation has been encountered for ILs in confined geometry. In a QENS study of [bmim+][Tf2N−] confined in the 8.5 nm diameter pores of a mesoporous carbon matrix, Chathoth et al. actually observed an increase of the dynamics compared to the bulk liquid.[27,28] They explained this behavior as a result of the higher density of ions adsorbed onto the pore wall, which leaves fewer, less densely packed ions, and thus faster dynamics, in the pore’s central region. In contrast, the homogeneous distribution of ethaline throughout the pore prevents such an effect from influencing its dynamics.

The comparison of the two isotopic compositions provides another interesting insight into the DES properties. Indeed, the impact of confinement on the dynamics is similar for both mixtures. As a results, there is no significant isotopic effect in the local relaxation time $\tau_L$, residence time $\tau_0$, and diffusion coefficient $D_T$. Having in mind that changing the isotopic composition increases the contribution from the hydrogenated component of the DES mixture relative to the deuterated counterpart, this absence of isotopic effect on the QENS measurements indicates that both EG and ChCl have comparable dynamics at the nanometer-scale. This observation was already reported for the bulk ethaline, while differences in diffusivities between the two DES components were observed at the micrometer-scale by NMR. This was assigned to the tendency of EG of ChCl to form supramolecular assemblies that are stabilized by strong electrostatic and H-bond interactions. At the nanometer-scale, the dynamics of both DES constituents are therefore highly coupled. Our findings demonstrate that this phenomenon of nanoscopic strong association remains true inside the DES mixture in confined geometry. The dynamics of both constituents are actually reduced by confinement but they remain highly coupled. A different situation appears in the bulk state at time exceeding the lifetime of this association. In this case, the diffusivities that was measured by NMR on the micrometer-scale obey the Stokes-Einstein law and vary according to the different hydrodynamic radii of EG and ChCl. This situation cannot be observed in nanoconfined geometry, because displacements larger than about 100 nm (typical porous silica grain size) are inhibited as long as the DES molecules remain inside the pore.

## CONCLUDING REMARKS

Currently, the development of hybrid nanomaterials incorporating DES confined within nanoporous geometries is generating significant interest across a broad spectrum of applied technologies.[7–13] In certain cases, this confinement can be strategically exploited to enhance the performance of deep eutectic solvents, such as extending the temperature range of the liquid-phase stability,[17,62] or improving gas absorption capabilities.[7,8]

Unlike conventional solvents, the performances of DESs are largely attributed to the specific interactions between their constituents, which give rise to long-lived organized structures, such as stoichiometric supramolecular ionic complexes.[49–52] The distinctive structural organization of DESs also significantly influences their dynamic behavior.[31,33–37,59,63–65] There is therefore a critical need to assess the impact of confinement on the structural organization of deep eutectic solvents (DESs). Regarding dynamics, a key question is whether confinement causes significant slowdowns, similar to those observed in many conventional liquids,[18,24,25,60,61] which could hinder applications requiring high ionic transport and low viscosity.

Our diffraction results show that ethaline remains well mixed inside the pores, without evidence for a core-shell segregation or demixing within the pore radius. In other words, confinement does not create a microphase separated liquid structure for this DES, in contrast with some other types of binary mixtures confined in the same porous hosts.[21,22,29,47] On the dynamical side, the combined time-of-flight and backscattering datasets consistently point to moderate confinement effects. Translational diffusion remains nearly as efficient as in the bulk, with only a small reduction of diffusivity (less than 20%) in the smaller pores. The key confinement signature concerns the localized motions: the residence time $\tau_0$ between two jumps, associated to the jump-diffusion mechanism, is longer by roughly a factor of 3-8 under confinement in SBA-15, and up to a factor of 10 in MCM-41 relative to bulk. Similarly, the localized in-cage motion that precedes diffusion is slowed down. The characteristic relaxation time, $\tau_L$, increases by approximately 20% for SBA-15 ($\tau_L \approx$ 3.5-4 ps) and by up to 50% for MCM-41 ($\tau_L \approx$ 4-6 ps). The associated trajectories, modeled from the EISFs in bulk DES,[31] are largely preserved under confinement. Only a slight reduction in motional amplitudes is observed, indicating that DES molecules experience marginally increased spatial constraints due to interactions with their neighboring molecules.

Overall, these structural and dynamical observations suggest that ethaline largely preserves its intrinsic bulk properties even when confined in nanopores. This conclusion holds significant promise for developing hybrid nanomaterials that incorporate DES, with potential to positively impact a wide range of technological applications. From this perspective, it is important to note that properties of DES are intrinsically tied to the interactions between its constituents, and, in confined geometry, to their interactions with the pore surface. Indeed, their response to confinement depends on the balance between surface affinity and the internal supramolecular network. As such, these properties highly depend on the chemical nature of both the DES and the porous host. Consequently, while our findings apply to ethaline confined in silica pores, they cannot be generalized to all DESs and porous materials, but rather underline the necessity of case-by-case studies of different classes of systems.

## SUPPLEMENTARY MATERIAL

See the supplementary material for the sample's neutron scattering cross sections; the theoretical modelling of the form factor of empty matrices and filled MCM-41; the information related to IN5B measurements (modelling, total scattering intensity, quasielastic broadening and EISF of the localized motions); the information related to IN16B measurements for DES confined in MCM-41 (elastic structure factor, EFWS; IFWS, translational diffusion); parameters obtained by modelling of QENS data.

## AUTHOR DECLARATIONS

### Conflict of Interest

The authors have no conflicts to disclose.

### Author Contributions

**Mohammad Nadim Kamar:** Data Curation (equal); Formal analysis (lead); Investigation (equal); Methodology (equal); Visualization (lead); Writing – original draft (lead); Writing – review & editing (equal). **Armin Mozhdehei:** Investigation (equal); Writing – review & editing (equal). **Ronan Lefort:** Investigation (equal); Writing – review & editing (equal). **Alain Moréac:** Investigation (equal); Writing – review & editing (equal). **Jacques Ollivier:** Data Curation (equal); Investigation (equal); Software (equal); Writing – review & editing (equal). **Markus Appel:** Data Curation (equal); Investigation (equal); Software (equal); Writing – review & editing (equal). **Viviana Cristiglio:** Data Curation (equal); Investigation (equal); Software (equal); Writing – review & editing (equal). **Denis Morineau:** Conceptualization (lead); Funding acquisition (lead); Investigation (equal); Methodology (equal); Project administration (lead); Resources (equal); Supervision (lead); Visualization (equal); Writing – original draft (equal); Writing – review & editing (equal).

## ACKNOWLEDGEMENTS

The experiments were conducted as part of the Ph.D. project of M. N. Kamar, who gratefully acknowledges funding from the University of Rennes. We express our gratitude to the Institut Laue-Langevin (ILL) for providing neutron beam time, which was central in the success of this study.

## DATA AVAILABILITY

The raw data generated at the Institut Laue-Langevin (ILL) large scale facility and the dataset will be publicly available after the embargo period (10.5291/ILL-DATA.6-07-112).[42] Derived data supporting the findings of this study are available from the corresponding author upon reasonable request.